\documentclass[12pt]{article}

\usepackage{amsmath,graphicx,bm,a4wide}

\newcommand{\super}[1]{$^{#1}$}
\newcommand{\average}[1]{\langle #1\rangle}
\newcommand{\addressstyle}{\em\normalsize}

\begin{document}

\title{Short-time fluctuations of~displacements~and~work}

\author{Ramses van Zon\super{\dagger*},	S. S. Ashwin\super{\ddagger*} and 
E. G. D. Cohen\super{*}
\\
\super{\dagger}\addressstyle 
Chemical Physics Theory Group, Department of Chemistry, University of Toronto,\\
\addressstyle
80 Saint George Street, Toronto, Ontario M5S 1A1, Canada\\
\super{\ddagger}\addressstyle
Department of Chemistry, University of Saskatchewan, \\
\addressstyle
110 Science Place, Saskatoon, Saskatchewan S7N 5C9, Canada\\
\super{*}\addressstyle
The Rockefeller University, 1280 York Avenue,  New York,\\
\addressstyle
New York 10021-6399, USA}

\date{\normalsize November 21, 2006}

\maketitle

\begin{abstract}
A recent theorem giving the initial behavior of very short-time
fluctuations of particle displacements in classical many-body systems
is discussed. It has applications to equilibrium and non-equilibrium
systems, one of which is a series expansion of the distribution of
work fluctuations around a Gaussian function. To determine the
time-scale at which this series expansion is valid, we present
preliminary numerical results for a Lennard-Jones fluid. These results
suggest that the series expansion converges up to time scales on the
order of a picosecond, below which a simple Gaussian function for the
distribution of the displacements can be used. \\


\end{abstract}

\section{Introduction}

In recent years, non-equilibrium fluctuations in entropy production,
work and heat have become the subject of intense study. In this
context, results such as the stationary state fluctuation
theorem\cite{FT} have given a deeper insight in the behavior and
character of non-equilibrium systems beyond irreversible
thermodynamics and hydrodynamics, which are especially important for
small systems such as biomolecules and nanotechnology.

The fluctuation theorem of a non-equilibrium steady state considers
fluctuations at long time-scales. Fluctuations of work, heat and
entropy production at \emph{short} time-scales are however less often
studied. Such fluctuations play a role in short-time (picosecond)
non-equilibrium transport at the nanometer scale such as in incoherent
neutron scattering, among others.  While fluctuations of many
properties can be studied, the present paper will give an overview of
recent results for short-time fluctuations of the displacements of
individual particles in a classical-mechanical system.  In particular,
we will focus on a theorem regarding the probability distribution of
the short-time fluctuations of
displacements\cite{VanZonCohen06b,VanZonCohen06a}. As we will show in
this paper, this distribution is also related to that of work
fluctuations. Furthermore, we will discuss its implications and range
of applicability.

The paper is structured as follows: In section
\ref{momentsandcumulants} we introduce moments and cumulants of a
probability distribution, in terms of which the theorem is
formulated. In section~\ref{example} we give a simple example of the
short-time behavior of cumulants as a prelude to the theorem presented
in section~\ref{theorem}.  One equilibrium and one non-equilibrium
application of this theorem are discussed in
section~\ref{applications}. Preliminary results on the range of
validity of the short-time expansion and thus of the utility of the
theorem are presented in section~\ref{numerics}. We end with the
conclusions in section~\ref{conclusions}.

\section{Moments and cumulants}
\label{momentsandcumulants}

Consider a classical system of $N$ particles with positions
$\bm{r}_{i}$ and velocities $\bm{v}_{i}$. The equations of motions are
Hamiltonian with forces derived from a smooth potential $U(\bm{r}_{1},
\ldots\bm{r}_{N})$, while the initial phase points have a probability
distribution denoted by $\rho (\{\bm{r}_{i}, \bm{v}_{i}\})$.  In the
current context, Lennard-Jones potentials may be considered smooth,
despite the singularity at zero, provided initial conditions which hit
this singularity have measure zero.

In time, the particles will be displaced from their initial position
by an amount $\Delta\bm{r}_{i}(t)=\bm{r}_{i}(t)-\bm{r}_{i}(0)=(\Delta
x_{i}(t),\Delta y_{i}(t),\Delta z_{i}(t))$. The probability
distribution of these displacements, $P (\Delta\bm{r}_{1}, \ldots
\Delta\bm{r}_{N},t)$, is determined by the dynamics and the initial
probability distribution.

Instead of considering the probability distribution function
$P(\Delta\bm{r}_{1}, \ldots \Delta\bm{r}_{N},t)$ itself, it is often
useful to consider the moments of this distribution which are defined
as
\begin{equation}
  \mu_{\bm{n}_{1}\ldots \bm{n}_{N}}(t)
= \average{\Delta x_{1}^{n_{1x}}\Delta y_{1}^{n_{1y}}\Delta z_{1}^{n_{1z}}
  \cdots 
\Delta x_{N}^{n_{Nx}}\Delta y_{N}^{n_{Ny}}\Delta z_{N}^{n_{Nz}}}_{t},
\end{equation}
where the average $\average{\,}_{t}$ is taken with
$P(\Delta\bm{r}_{1}, \ldots\Delta\bm{r}_{N},t)$ and each $\bm{n}_{i}$
is triplet of integers, $\bm{n}_{i}=(n_{ix},n_{iy},n_{iz})$. These
moments determine the moment generating function
\begin{equation}
    \hat{P}(\bm{k}_{1}, \ldots \bm{k}_{N},t) =
    \sum_{n_{1x}=0}^{\infty}\frac{(ik_{1x})^{n_{1x}}}{n_{1x}!}
    \sum_{n_{1y}=0}^{\infty}\frac{(ik_{1y})^{n_{1y}}}{n_{1y}!}
    \sum_{n_{1z}=0}^{\infty}\frac{(ik_{1z})^{n_{1z}}}{n_{1z}!}
    \cdots
    \sum_{n_{Nz}=0}^{\infty}\frac{(ik_{Nz})^{n_{Nz}}}{n_{Nz}!}
\: \mu_{\bm{n}_{1}\ldots\bm{n}_{N}}(t),
\label{Phatdef}
\end{equation}
which coincides with the Fourier transform of
$P(\Delta\bm{r}_{1},\ldots\Delta\bm{r}_{N},t)$.  Thus all information
about $P(\Delta\bm{r}_{1},\ldots\Delta\bm{r}_{N},t)$ is in principle
contained in these moments.

A special case is formed by taking $\bm{n}_{1}=(n,0,0)$ and all other
$\bm{n}_{i\neq1}=0$, which means one considers the distribution of the
displacement of a single particle, $P(\Delta x,t)$.  Its moments are
given by $\mu_{n}(t)=\average{\Delta x^{n}_{1}}_{t}$ and its
generating function by $\hat{P}(k,t) = \sum_{n=0}^{\infty} \mu_{n}(t)
= \average{\exp(ik\Delta x)}_{t}$. In equilibrium, $\hat{P}(k,t)$
becomes equal to the incoherent intermediate scattering function that
can be measured e.g. by neutron scattering.

In special cases, such as for an ideal gas or a harmonic lattice in
equilibrium, the moment generating function is Gaussian and then
determined by the first and second moments, i.e, for the single
particle displacement by the $\mu_{1}(t)$ and $\mu_{2}$$(t)$. Even in
those cases, the higher order moments are not zero, but are simple
factored forms involving $\mu_{1}$ and $\mu_{2}$.

A more convenient representation of the distribution for Gaussian and
near-Gaussian distributions is in terms of cumulants, which are,
loosely stated, moments with all possible factorizations taken
out. Formally, they are defined as the derivatives of the logarithm of
$\hat{P}$, such that:
\begin{equation}
  \sum_{n=1}^{\infty} \frac{(ik)^{n}}{n!}\kappa_{n}(t)=\log\hat{P}(k,t).
\label{kappadef}
\end{equation}
This equation defines the cumulants $\kappa_{n}(t)$ for the single
particle displacement distribution $P(\Delta x,t)$, but it can easily
be generalized to deal with the general case of
$P(\Delta\bm{r}_{1},\ldots\Delta\bm{r}_{N},t)$.  Note that in case of
a Gaussian distribution, the cumulants $\kappa_{n}$ with $n>2$ are all
equal to zero.

One can find explicit expressions for the cumulants by substituting
the expression on the right-hand side of equation~\eqref{Phatdef} into
the right hand side of equation~\eqref{kappadef}, expanding the
logarithm and equating terms of equal powers in $k$. For the first
four powers of $k$ this yields
\begin{subequations}%
\begin{eqnarray}
  \kappa_{1}(t) &=& \mu_{1}(t)\\
  \kappa_{2}(t) &=& \mu_{2}(t)-\mu_{1}^{2}(t)\\
  \kappa_{3}(t) &=& \mu_{3}(t)-3\mu_{1}(t)\mu_{2}(t)+2\mu_{1}^{3}(t)\\
  \kappa_{4}(t) &=& \mu_{4}(t)-4\mu_{1}(t)\mu_{3}(t)-3\mu_{2}^{2}(t)
                    +12\mu_{1}^{2}(t)\mu_{2}(t)-6\mu_{1}^{4}(t).
\end{eqnarray}%
\end{subequations}%
These first four cumulants are also known as the mean, variance,
skewness and kurtosis, respectively.

\section{Short-time behavior: a harmonic example}
\label{example}

As an illustrative example, consider a single particle of mass one in
one dimension, subject to a harmonic potential $U=x^{2}/2$. Let the
initial velocity distribution be Gaussian:
\begin{equation}
  \rho(v_{0}) = (2\pi k_{B}T)^{1/2}
\exp\left[-\frac{v_{0}^{2}}{2k_{B}T}\right].
\label{vdist}
\end{equation}
The distribution of the initial positions $x_{0}$ will be arbitrary
and given by some function $p(x_{0})$. Because $p$ is arbitrary, this
single-particle system is not necessarily in equilibrium.

The equations of motion of this system are easily solved and give 
\begin{equation}
  x(t)= \cos t \,x_{0} + \sin t \, v_{0}.
\label{dynamics}
\end{equation}
Using equations \eqref{vdist} and \eqref{dynamics}, the function
$\hat{P}(k,t)$, i.e., the Fourier transform of $P(\Delta x, t)$ with
$\Delta x=x(t)-x_{0}$, can now be computed:
\begin{eqnarray}
  \hat{P}(k,t) &=&
  \int\!\mathrm{d}x_{0}\int\!\mathrm{d}v_{0}\:  p(x_{0}) \rho(v_{0}) e^{ik[x(t)-x_{0}]}
\nonumber\\&=&
  \int\!\mathrm{d}x_{0}\int\!\mathrm{d}v_{0}\:  p(x_{0}) \rho(v_{0}) 
  e^{ik[(\cos t-1)x_{0}+\sin t \,v_{0}]}
\nonumber\\&=&
  \int\!\mathrm{d}x_{0}\: p(x_{0}) e^{ik(\cos t-1)x_{0}}
  \int\!\mathrm{d}v_{0}\: \rho(v_{0})  e^{ik\sin t\, v_{0}}
\nonumber\\&=&
\hat{p}(k[\cos t -1])\,
\exp\left[-\frac{k^{2}}{2}k_{B}T\sin^{2} t\right],
\end{eqnarray}
where $\hat{p}$ is the Fourier transform of the distribution of
initial positions, $p(x_{0})$.  Given this result for $\hat{P}(k,t)$
and equation~(\ref{kappadef}), one finds for the cumulants:
\begin{equation}
  \kappa_{n}(t)=\kappa_{n}^{(0)}(\cos t-1)^{n}+\delta_{n,2}k_{B}T\sin^{2} t,
\label{kappaexample}
\end{equation}
where $\kappa_{n}^{(0)}$ is the $n$th cumulant of $p(x_{0})$.

\begin{figure}[t]
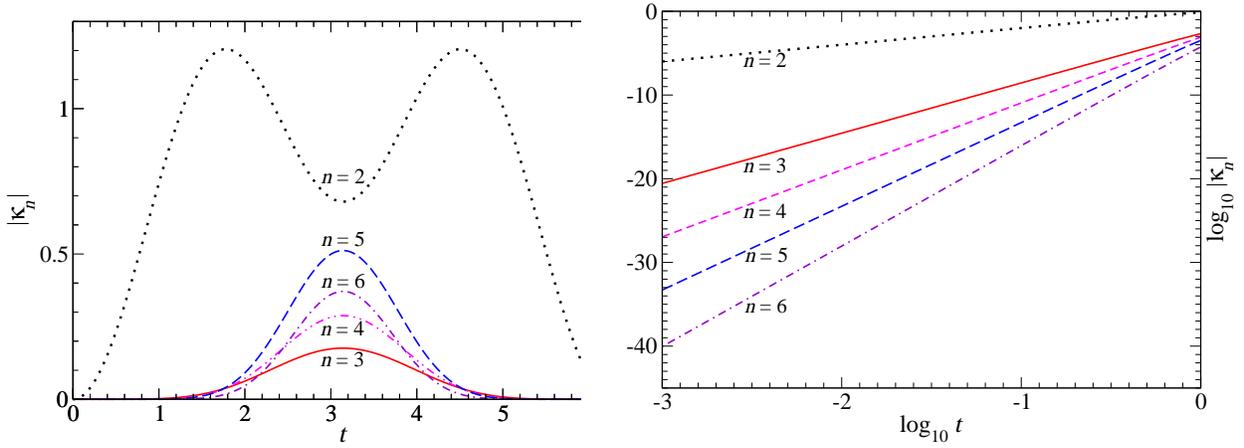

\centerline{
\includegraphics[width=0.5\textwidth]{figure1a}
\includegraphics[width=0.5\textwidth]{figure1b}}
\caption{Short-time behavior of cumulants of the displacement of a
particle in a harmonic potential, initially not in equilibrium, based
on equation (8).  The left plot shows the cumulants $\kappa_{n}$ as a
function of time $t$, for $n=2$, 3, 4, 5 and $6$, respectively. The
right plot shows the same on a log-log scale, to bring out the scaling
behavior for small~$t$.}\label{examplefig}
\end{figure}

An example of the cumulant behavior in equation (\ref{kappaexample})
is given in figure~\ref{examplefig}. For this example, we chose
$k_{B}T=1$ and for $p(x_{0})$ a non-Gaussian function which is
continuous and piece\mbox-wise linear:
\begin{equation*}
p(x_{0}) = \begin{cases}
  \frac{3}{4}(1+x_{0}) &\text{ if }-1<x_{0}<\frac{1}{3}\\
  \frac{3}{2}(1-x_{0}) &\text{ if }\quad\frac{1}{3}<x_{0}<1\\
  0&\text{ otherwise.}
	 \end{cases}
\end{equation*}
The reason for choosing this $p(x_{0})$ is that its cumulants are
non-trivial yet easy to compute.  The first panel in
figure~\ref{examplefig} shows the behavior of the cumulants
$\kappa_{n}$ for $n=2$, 3, 4, 5 and 6 as a function of time. The
second panel shows the same quantities, but zooms in on the small time
behavior and has a log-log scale. On this log-log scale, the curves
for $n=2$, 3, 4, 5 and 6 have, for small $t$, slopes $-2$, $-6$, $-8$,
$-10$ and $-12$, respectively. This means that for small $t$,
$\kappa_{2}\propto t^{2}$, while $\kappa_{2<n\leq6}\propto t^{2n}$. It
can easily be verified that equation~(\ref{kappaexample}) supports
this observation for all choices of $p(x_{0})$ and for all $n>2$.

One may wonder whether the $t^{2n}$ behavior of $\kappa_{n}(t)$ for
small times $t$ is general, or whether it is due to the simplicity of
our example. In fact, this behavior is more general, under conditions
explained in the next section.

\section{Theorem on short-time fluctuations of displacements}
\label{theorem}

Cumulants of particle displacements have been studied in the context
of the intermediate scattering function in the field of neutron
scattering on equilibrium systems for a long
time\cite{Schofield61,Sears72,DeSchepperetal81}. It was found that
while all odd cumulants of displacements are zero in equilibrium, the
even cumulants for small times obey $\kappa_{n}(t)\propto |t|^{n+1}$
for hard spheres\cite{DeSchepperetal81}, while for systems with smooth
potentials it was known that $\kappa_{4}(t)\propto t^{8}$
\cite{Schofield61} and $\kappa_{6}(t)\propto t^{12}$ \cite{Sears72}.

Unfortunately, these results for smooth potentials only held for
equilibrium systems and furthermore no result s for general $n$ were
available. A recent theorem however, generalizes these results to
certain non-equilibrium systems and to general
$n$\cite{VanZonCohen06b,VanZonCohen06a}. It states that if
\begin{enumerate}
\item the forces are smooth functions of the positions of the
  particles (smooth here means that all derivatives exist for all
  points in phase space except a set of measure zero),
\item the forces are independent of the velocities,
\item the initial velocities are Gaussian distributed, and
\item the initial velocities are independent of the initial positions,
\end{enumerate}
then the cumulants of particle displacements satisfy for small times
$t$
\begin{equation}
   \kappa_{n}(t) = \left\{\begin{array}{ll}
c_{n} t^{n} + \mathcal{O}(t^{n+1})&\mbox{for }n<3
\\
c_{n} t^{2n} + \mathcal{O}(t^{2n+1})&\mbox{for }n\geq 3,
			\end{array}
\right.
\label{theoremt}
\end{equation}
where the $c_{n}$ are constants. In fact, all $n$-order cumulants
satisfy this rule (an $n$-order cumulant is cumulant
$\kappa_{\bm{n}_{1}\ldots\bm{n}_{N}}$ such that $\sum_{i}
(n_{ix}+n_{iy}+n_{iz}) = n$).

We stress that this theorem applies to non-equilibrium as well as to
equilibrium systems, because the (initial) position distribution can
be arbitrary and does not have to be equal to its equilibrium form.  A
common class of non-equilibrium systems covered by the theorem
consists of those that start out in equilibrium and are then driven
out of equilibrium by some external perturbation, because the
requirement of a Gaussian velocity distribution only has to be
satisfied at the initial time.

A sketch of the proof of this theorem was given in reference~2, while
the full proof can be found in reference~3; we will therefore not
prove the theorem here again. However, we do want to remark that the
nature of the dynamics is largely unimportant for the theorem to hold
(as long as the forces are smooth), and that the most important
requirement seems to be that the velocities are initially Gaussian
distributed.  To what extent one can relax the requirement that this
Gaussian distribution be independent of the initial positions, is at
present not clear.

\section{Applications}
\label{applications}

\subsection{Neutron scattering in equilibrium systems}
As mentioned above, the short time behavior of the cumulants of
displacement arises in the context of neutron scattering. In fact, in
equilibrium, the incoherent structure function $F_{s}(k,t)$ is
precisely the Fourier transform of the distribution of single particle
displacements $\hat{P}(k,t)$\cite{HansenMcDonald}. Given
equation~(\ref{kappadef}), we can therefore write
\begin{equation}
  F_{s}(k,t) 
= \exp\left[\sum_{n=2}^{\infty} \frac{\kappa_{n}(t)}{n!}(ik)^{n}\right].
\label{Fsexp}
\end{equation}
Note that all odd cumulants are zero for an equilibrium system 
for symmetry reasons.

For small wave-vector $k$, the terms involving higher cumulants in
equation~(\ref{Fsexp}) can be neglected and $F_{s}(k,t)$ becomes
approximately Gaussian, as does its Fourier inverse, the Van Hove
self-correlation function $G_{s}(\Delta x,t)$. While
equation~(\ref{Fsexp}) in principle allows for a systematic
improvement on this Gaussian approximation by increasing the number of
terms considered in the cumulant sum, in its current form, it is not
possible to take into account higher order cumulants and still perform
the Fourier inverse analytically. To obtain a systematic analytic
improvement upon the Gaussian approximation of the Van Hove
self-correlation function one first has to expand the right-hand side
of equation~(\ref{Fsexp}) around a Gaussian as follows:
\begin{equation}
  F_{s}(k,t) = \exp\left[-\frac{1}{2}\kappa_{2}(t) k^{2}\right]
\left[ 1 + \frac{\kappa_{4}(t)}{4!}k^{4}-\frac{\kappa_{6}(t)}{6!}k^{6}+\ldots\right].
\label{Fsexpanded}
\end{equation}
By taking the Fourier inverse of this expression, we find for the Van
Hove self-correlation function
\begin{equation}
 G_{s}(\Delta x,t)=\frac{\exp(-w^{2})}{\sqrt{2\pi\kappa_{2}(t)}}\left[1 +
  \frac{\kappa_{4}(t)H_{4}(w)}{4!4\kappa_{2}^{2}(t)} +
  \frac{\kappa_{6}(t)H_{6}(w)}{6!8\kappa_{2}^{3}(t)}+\cdots \right],
\label{Gsexpanded}
\end{equation}
where $w=\Delta x/\sqrt{2\kappa_{2}}$ and $H_{n}$ is the $n$th Hermite
polynomial.

Provided that the system has smooth forces independent of the
velocities (conditions 1 and~2 of the theorem, respectively), the
third and fourth conditions are automatically satisfied in (canonical)
equilibrium.  The application of the theorem now consists of noting
that, with $w=\mathcal{O}(1)$, the terms inside the brackets on the
right-hand side of equation~(\ref{Gsexpanded}) are $1$, $t^{4}$,
$t^{6}$, respectively. It can be shown that this behavior persists,
i.e., each next term is two powers of $t$ higher than the previous
one. Hence as a consequence of the theorem, the subsequent terms in
this expansion become successively smaller if $t$ is small, so that a
truncation of the expansion can give an accurate approximation to
$G_{s}(\Delta x,t)$.

\subsection{Distribution of work done by an external field}

Since the theorem is not restricted to equilibrium systems, we want to
discuss at least one interesting application of the theorem to a
non-equilibrium system, namely a system in a constant external field
along the $x$ direction.  This constant external field exerts a
constant force $\bm{F}$ on all $N$ particles, which does an amount of
work $W$ on the system. Since the initial conditions are drawn from a
probability distribution, $W$ will be a fluctuating
quantity. Non-equilibrium fluctuations of work have been investigated
in the context of the fluctuation theorem\cite{FT} and the
non-equilibrium work relation\cite{Jarzynski}. The theorem of
section~\ref{theorem} can be used to investigate a different aspect of
the work fluctuations, namely their very short-time behavior.

Consider now the work $W$ done on the system by this force during a
time $t$:
\begin{eqnarray*}
  W &=& \sum_{i=1}^{N} \int_{0}^{t}\: \mathrm{d}t'\:\bm{F}\cdot\bm{v}_{i}(t')
\nonumber\\&=&
|\bm{F}|\sum_{i=1}^{N} \Delta x_{i}(t).
\end{eqnarray*}
Given this linear relation between the displacements and the work, the
short-time theorem can be applied to the distribution of work values
as well. To see this, one can apply the multinomial
expansion\cite{VanZonCohen06a} to the $n$th cumulants of the work
distribution, i.e.
\begin{equation}
  \kappa_{n}^{W} (t) = \sum_{\sum_{j=1}^{N}n_{j}=n}\frac{n!}{n_{1}!n_{2}!\dot{s} n_{N}!}
\kappa_{n_{1}n_{2}\dot{s} n_{N}}(t),
\label{kappaw}
\end{equation}
where $\kappa_{n_{1}n_{2}\dot{s}
n_{N}}=\kappa_{\bm{n}_{1}\bm{n}_{2}\ldots\bm{n}_{N}}$ with
$\bm{n}_{j}=(n_{j},0,0)$.  Examples are $\kappa_{1}^{W} =
N\kappa_{100\ldots}$ and $\kappa_{2}^{W} = N\kappa_{200\ldots} +
N(N-1)\kappa_{1100\ldots}$ (where we used that the particle are
indistinguishable). If the forces are smooth and independent of the
particles' velocities and if the initial velocities of the particles
are Gaussian distributed, then according to the theorem, each term in
the sum on the right hand side of equation (\ref{kappaw}) is of order
$t^{2n}$ if $n>2$.  In other words, the $n$th cumulant of the work
fluctuations $\kappa^{W}_{n}(t)$ also scales as $t^{2n}$ for $n>2$ if
$t$ is small enough. Consequently, the distribution of work
fluctuations may be written in a similar expansion as that of the Van
Hove self-correlation function in equation~(\ref{Gsexpanded}).

\pagebreak[3]
\section{Relevant time scale for an equilibrium Lennard-Jones fluid}
\label{numerics}

We have discussed the short-time behavior of the cumulants of
displacements, but we have not yet addressed the question of how short
the time scale has to be for the above applications to be meaningful.
We will now address this question now for single particle
displacements, which play a significant role in the Van Hove
self-correlation function $G_{s}$.  We would call the series expansion
of $G_{s}$ in equation~(\ref{Gsexpanded}) meaningful if each next term
is smaller than the previous one. While the theorem guarantees that
this is so below some time-scale, it does not say what this time scale
is. An estimate for the time scale at which these short-time results
will no longer be applicable could be found by considering the time
$t=\tau_{G}$ at which the first correction term in
equation~(\ref{Gsexpanded}) becomes of comparable magnitude as the
first term (the Gaussian). In that case, the series in
equation~(\ref{Gsexpanded}) can no longer be expected to be useful.
Note that in equation~(\ref{Gsexpanded}) all odd cumulants have
vanished because of time-reversal symmetry, so the first correction
term involves $\kappa_{2}$ and $\kappa_{4}$, or, using
equation~(\ref{theoremt}), the values of the constants $c_{2}$ and
$c_{4}$. Since $w=\mathcal{O}(1)$, one can find $\tau_{G}$ by equating
the first and (the absolute value of) the second term in
equation~(\ref{Gsexpanded}) with $t$ set to $\tau_{G}$ and $H(w)$ set
to~$1$:
\begin{equation*}
  1 =   \left|\frac{c_{4}\tau_{G}^{8}}{4!4[c_{2}\tau_{G}^{2}]^{2}}\right|, 
\end{equation*}
where equation~(\ref{theoremt}) was used. Solving this equation for
$\tau_{G}$ yields
\begin{equation}
 \tau_{G} = \left(\frac{96 c_{2}^{2}}{|c_{4}|}\right)^{1/4}.
\label{tauG}
\end{equation}

Expression for $c_{n}$ can be found in section 3.4 of reference 3.
While the values of the $c_{n}$ will be system dependent, but it is
nonetheless of interest to have a realistic estimate of
$\tau_{G}$. For this purpose, we performed molecular dynamics (MD)
simulations at constant volume and energy (NVE) for an equilibrium $N
= 100$ particle system with periodic boundary
conditions\cite{FrenkelSmit}. The inter-atomic potential used is the
Lennard-Jones (LJ) potential
$V(r)=4\epsilon[(\sigma/r)^{12}-(\sigma/r)^{6}]$.  All quantities
reported are in Lennard-Jones units: length in units of $\sigma$,
temperature in units of $\varepsilon/k_{B}$, number density ($\rho$)
in units of $\sigma^{-3}$ and time in units of $\tau_{LJ}= (\sigma^{2}
m/ \epsilon)^{1/2}$, where $m$ is the mass of the particle.  Since
these are arbitrary units, to understand the physical consequences of
our results, we use the LJ parameters of a specific substance as a
reference, namely Argon. In that case, $\tau_{LJ}$ corresponds to
$2.16$ picoseconds while $\epsilon/k_{B} = 119.8$~K
\cite{FrenkelSmit}. In the simulation, a potential cutoff of $r_{c} =
2.5\sigma$ was used and the equation s of motion were integrated using
the Verlet algorithm\cite{FrenkelSmit} with a time step of one
femtosecond. Data were accumulated once equilibrium had been attained
in the simulation and collected between sufficiently long periodic
time intervals, in order to ensure statistically uncorrelated data
points.

Note that while strictly the theorem should be applied to a system in
the canonical (NVT) ensemble (to ensure a Gaussian velocity
distribution), the preliminary data presented here are obtained in the
microcanonical ensemble (NVE). Since in the thermodynamic limit, these
two ensemble should become equivalent, the time scale $\tau_{G}$
obtained in the NVE simulation is expected to be very close to the
$\tau_{G}$ that would be found in an NVT simulation.

\begin{figure}[tb]
\centerline{\includegraphics[width=0.65\textwidth]{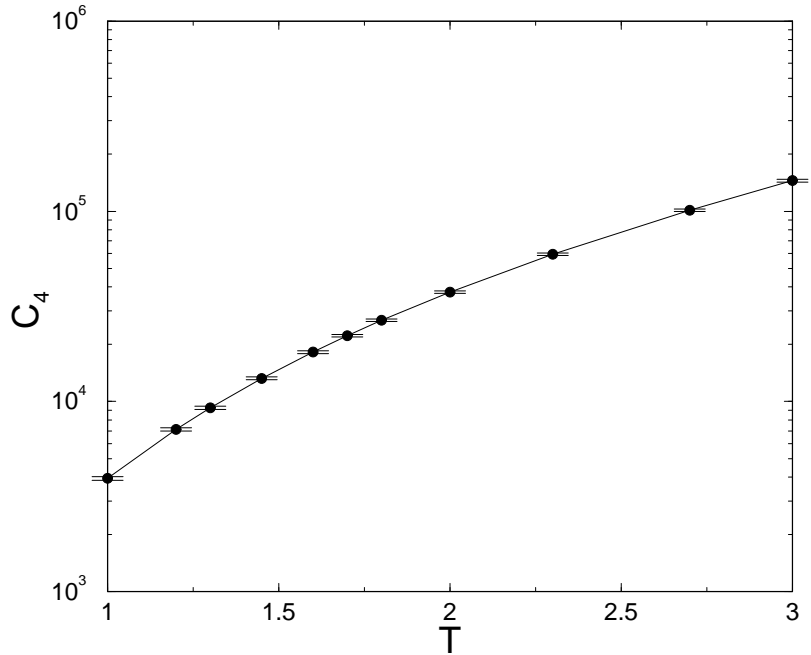}}
\caption{Numerical results for $c_{4}$, the coefficient in front of the
$t^{8}$ behavior of $\kappa_{4}$, as a function of the kinetic temperature
$T$.  The horizontal lines through the points are the error-bars.  The
system is a Lennard-Jones fluid with density $\rho=0.8$. These results
are from a MD simulation in an NVE ensemble with $N=100$ particles,
with periodic boundary conditions. All quantities are in LJ units.}
\label{C4}
\end{figure}

In figure~\ref{C4}, we show preliminary results from these simulations
for $c_{4}$ as a function of the kinetic temperature $T$.  More
results will appear in reference 10.  The value of $c_{2}=k_{B}T$, so
from the results in figure \ref{C4} we can estimate the time scale
$\tau_{G}$ using equation~(\ref{tauG}). It is found that this time
scale ranges from $\tau_{G}=0.3$ for $T=1$ to $\tau_{G}=3$ for $T=3$,
all in Lennard-Jones units. This corresponds to time scales on the
order of $0.6$ picoseconds for $T=119.8$~K to $6$ picoseconds for
$T=359.4$~K.  We note that these time scales are to be interpreted as
upper bounds on the times at which expansions such as those in
equation~(\ref{Gsexpanded}) break down.

\section{Conclusions}
\label{conclusions}

We discussed a recent theorem which states that for smooth systems
with initial Gaussian velocity distributions (but with arbitrary
position distributions), the cumulants satisfy a scaling relation for
short times: $\kappa_{n}\propto t^{n}$ if $n<3$ and $\kappa_{n}\propto
t^{2n}$ for $n\geq 3$. As applications of this theorem, this scaling
implies that the expansion of the Van Hove self-correlation function
around a Gaussian is useful, as is a similar expansion for the
short-time work fluctuations in a non-equilibrium system subject to a
constant external field (provided the initial distribution of
velocities is Gaussian).

We also showed some preliminary numerical results for a Lennard-Jones
model of Argon which suggest that this expansion breaks down for
larger times, with a n upper limit of $0.6$ to $6$ ps depending on the
kinetic temperature. Further details will be published in a later
paper\cite{Ashwinetal07}.

Finally, we wish to note that other applications of the theorem are
the scaling of non-Gaussian parameters characterizing ``dynamical
heterogeneities'' in supercooled liquids and glasses, and
non-equilibrium transport using a Green's functions approach; see
reference~3 for details.

\section*{Acknowledgments}

This work was supported by the Office of Basic Energy Sciences of the
US Department of Energy under grant number DE-FG-02-88-ER13847 and by
the National Science Foundation under grant number PHY-05011315.

\end{document}